\documentclass{iaus}
\usepackage{graphicx}

\title[Spitzer Spectra of Magellanic Cloud PNe] 
{The IRS Spitzer Spectra of the Magellanic Cloud Planetary Nebulae: Revealing the Dust and Gas Chemistry}

\author[Stanghellini et al.]   
{L.~Stanghellini$^{1,2}$,
P.~Garc\'\i a-Lario$^{3,4}$,
A.~Manchado$^5$,
J.~V.~Perea-Calder\'on$^3$,
D.~A.~Garc\'\i a-Hern\'andez$^3$,
R.~A.~Shaw$^1$,
E.~Villaver$^6$}

\affiliation{$^1$National Optical Astronomy Observatory, 950 N Cherry Avenue, Tucson AZ 85719, USA; $^2$ On leave, INAF-Bologna Observatory;
$^3$  European Space Astronomy Centre. Research and Scientific Support Department of ESA. Villafranca del Castillo, P.O. Box 50727. E-28080 Madrid, Spain;
$^4$ ISO Data Centre and Herschel Science Centre. European Space Astronomy Centre. Research and Scientific Support Department of ESA. $^5$ Instituto de Astrof\'isica de Canarias, v\'{\i}a L\'actea s/n, La Laguna, E-38200 Tenerife, Spain; affiliated to CSIC, Spain;
$^6$ Space Telescope Science Institute, 3700 San Martin Drive, Baltimore MD 21218, USA; affiliated to the Hubble Space Telescope Department of ESA

}

\pubyear{2006}
\volume{234}  
\pagerange{}
\date{?? and in revised form ??}
\jname{Planetary Nebulae in our Galaxy and Beyond}
\editors{M. J. Barlow \& R. H. M\'endez, eds.}
\begin{document}

\maketitle

\begin{abstract}

Planetary nebulae (PNe) in the Magellanic Clouds (LMC, SMC) offer a unique opportunity to study both the 
population and evolution of low- and intermediate-mass stars in an environment which is free of the 
distance scale bias that hinder Galactic PN studies.  The emission shown by PNe in the 5-40 $\mu$m
 range is characterized by the presence of a combination of solid state features (from the dust grains) 
 and nebular emission lines over-imposed on a strong dust continuum. We acquired low resolution IRS 
 spectroscopy of a selected sample of LMC and SMC PNe whose morphology, size, central star brightness, 
 and chemical composition are known. The data have been acquired and reduced, and the IRS spectra show 
 outstanding quality as well as very interesting features. 
The preliminary analysis presented here allows to determine strong correlations between gas and dust composition, 
and nebular morphology. More detailed analysis in the future will deepen our knowledge 
of mass-loss mechanism, its efficiency, and its relation to PN morphology.

\end{abstract}

\begin{figure}
 \includegraphics[width=11truecm]{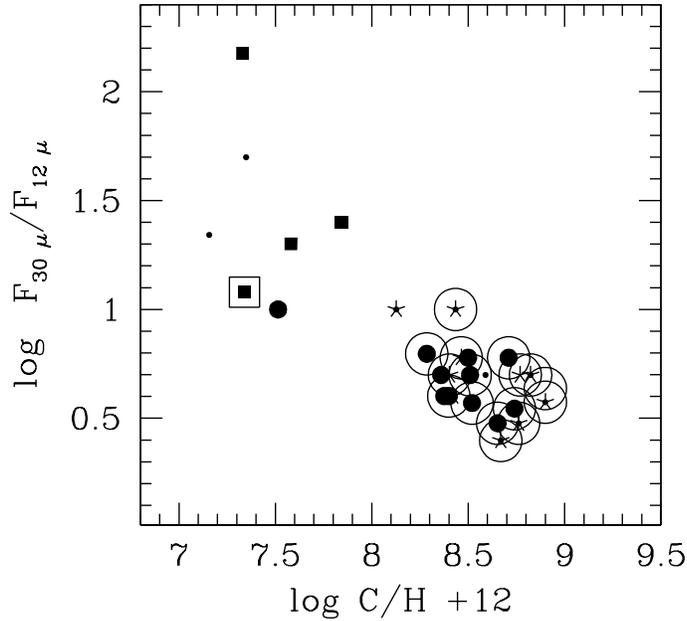}
  \caption[]{The dust production efficiency, measured from the ratio of the 
  nebular continuum at 30 and 12 $\mu$m, versus the carbon abundance of the gaseous component 
  (Stanghellini et al. 2005) in unresolved (tiny circles),
  round (filled circles), elliptical (asterisks),
  and bipolar or quadrupolar (filled squares) Magellanic Cloud PNe. Large open circles: C-dust detection, large 
  open square: O-dust detection.}
\end{figure}

\section{Motivation}
The transition between the Asymptotic Giant Branch (AGB) and planetary nebula (PN) phases is controversial. 
The occurrence of the major mass-loss that stars undergo at the tip of the AGB, its magnitude and duration, 
its relation to the  metallicity of the progenitor and the environment, and the post-ejection evolution of both 
stars and nebulae, are still only poorly constrained by observations. Open issues include the mechanism of mass-loss,
its relation to metallicity and to progenitor mass, the coupling of the dust and gas component, and the 
dependence of morphology on metallicity and dust content of PNe. 

From the theoretical point of view, if mass-loss at the AGB tip occurs by radiation pressure
on the dust grains, then its efficiency increases with metallicity (Willson 2000). On the other hand,
mass-loss may occur also, with lower efficiency, in the absence of dust grains. It is worth recalling that not only the envelope ejection
phase, but also the post-AGB evolution, is affected by the nature of dust, and that the 
{\it thinning} time of a PN at a given wavelength is proportional to the mass-loss rate and to the absorption coefficient
of the grain material (Kaeufl et al. 1993).

Observations of AGB and post-AGB stars in various environments seem to indicate that metallicity plays a
fundamental role in AGB mass-loss. It is observed, in fact, that the number of highly obscured AGB stars
is smaller in the Magellanic Clouds that in the Galaxy (Groenewegen et al. 2000; Trams et al. 1999), indicating that
a low metallicity environment is compatible with lower mass-loss efficiency. It is also known that the
ratio of carbon-rich to oxygen-rich AGB stars decreases with metallicity (Cioni \& Habing 2003). Finally,
from the analysis of PN morphology in the SMC and the LMC, Stanghellini et al. (2003) found that there are
more deviations from spherical symmetry in the relatively higher metal environment of the LMC than in the SMC.

From these premises it seemed that an accurate study of the correlation between the PN dust features and
the gas metallicity was needed to link the dust properties to PN morphology, and
to determine the evolution of the dust features. We decided to perform such an investigation in 
the Magellanic Clouds, in order to acquire dust features for a PN sample that was (1) free from distance scale
biases, (2) had easy access trough a low-reddening field, (3) had a wide metallicity domain, and
(4) had HST imaging counterparts, to eliminate the 
possible misclassified objects in the sample. 

\section{Observations and Preliminary Results}

We acquired IRS/Spitzer spectra
for all compact (with radius less than 2 arcsec) confirmed Magellanic Cloud PNe within the reach of Spitzer, 
and that were not in the Reserved Object Catalog of Spitzer. Our low resolution IRS spectra allow
a good estimate of the dust continuum temperature, and the study of the solid state features and 
nebular emission lines.

At the time of writing, we have completed a preliminary analysis of the spectra: we have checked the pointing quality, 
performed the
cleanup of the 2D spectra, and extracted the 1D spectrograms. We find that the few dozen PNe observed display a 
variety of continuum strengths and dust temperatures. Superimposed to the continua we see the nebular forbidden
lines of [Si VII], [Ne IV], [Si IV], [Ne III], [Ne V], and [O IV]. Furthermore, we could identify
the C-rich dust features relative to the PAH series, and to what it seems to be SiC broad emission. Finally, O-rich crystalline
silicates have been identified in one target.

In Figure 1 we plot the ratio of the IRS continuum measured at 30 and 12 $\mu$m against
the gaseous carbon abundance 
(from Stanghellini et al. 2005). The IR flux ratio plotted in Fig. 1 
is a measure of the dust production efficiency.
We plotted the morphological types with different symbols (see figure legend), 
and also added large open circles to PNe where carbon dust has been detected, and a large open square to the only 
PN where crystalline silicates  have been detected. The resulting plot shows a strong anti-correlation between dust
production efficiency and carbon abundance. Since carbon is depleted in the most massive AGB stars,
where the 
hot bottom burning (HBB) is activated, it appears that the dust production is more efficient where the HBB is
also efficient. Maximum dust production occurs for the most aspheric PNe, such as those with bipolar and quadrupolar
shape.

In Figure 2 we plot the gas composition of a partial sample of PN in the C/O and N/O plane, showing
the known morphological segregation due to the different evolutionary paths of spheric and aspherical
PN progenitors (see Stanghellini et al. 2005 and citation therein for abundances). 
In the same plane we also show the
findings from our Spitzer observations, i.e, large open circles and open squares corresponding to C-rich and
O-rich dust. We see that C-rich dust detection pertains to those PNe with gaseous C/O$>$1, all of them
round or elliptical in shape. On the other hand, all C/O$<$1 PNe are bipolar or quadrupolar, and the only O-rich dust
detection corresponds to  a nitrogen rich, carbon poor PN.

\begin{figure}
 \includegraphics[width=11truecm]{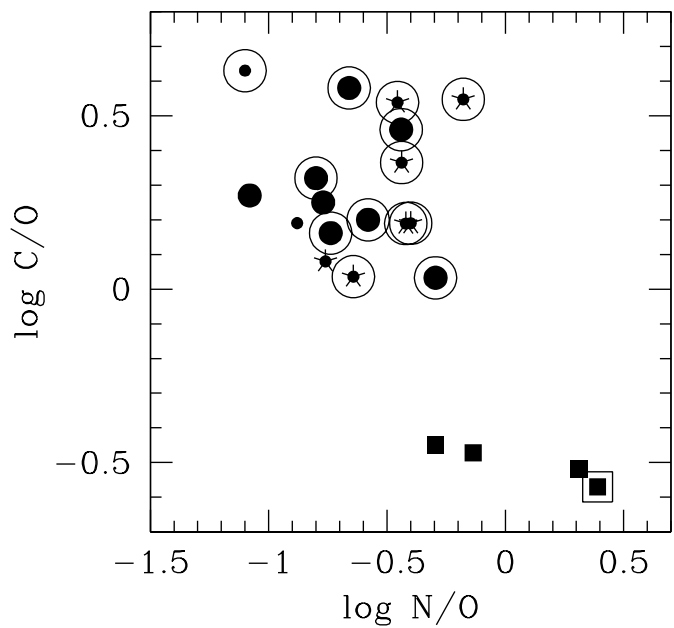}
  \caption[]{C/O plotted against N/O for the LMC and SMC PN of our Spitzer sample. 
  Gas abundances are from Stanghellini et al. 2005, Leisy \& Dennefeld 1996, and Stasinska et al. 1996.
  Symbols are as in Fig. 1.}
\end{figure}

We plan to refine the above analysis and extend it to the whole PN sample. Furthermore, we plan to complete the quantitative 
analysis of the dust features, including continuum
subtraction and measurement of the PAH and silicate features. Finally, a detailed study of the IR nebular emission
lines is in progress, including the modeling of these LMC and SMC PNe with photo-ionization codes, to match the
emission in the optical, IR, and UV ranges.


\begin{thebibliography}{}

\bibitem[]{}Cioni, M.-R.L. \& Habing H.J 2003. A\&A 402, 133
\bibitem[]{}Groenewegen, M.A.T., Blommaert, J.A.D.L., Cioni, M.-R.L. et al. 2000, MmSAI 71, 639
\bibitem[Kaeufl et al.(1993)]{1993ApJ...410..251K} {} Kaeufl, H.~U., Renzini, A., \& Stanghellini, L.\ 1993, ApJ, 410, 251 
\bibitem[Stanghellini et al.(2003)]{2003ApJ...596..997S} Stanghellini, L., 
Shaw, R.~A., Balick, B., Mutchler, M., Blades, J.~C., \& Villaver, E.\ 
2003, ApJ, 596, 997 
\bibitem[Stanghellini et al.(2005)]{2005ApJ...622..294S} Stanghellini, L., 
Shaw, R.~A., \& Gilmore, D.\ 2005, ApJ, 622, 294 
\bibitem[]{}Trams, N.R., van Loon, J.Th., Waters, L.B.F.M. et al. 1999, A\&A 346, 843
\bibitem[]{}Willson, L.~A.\ 2000, ARA\&A, 38, 573 

     


\end{thebibliography}
\end{document}